\documentclass[aps,twocolumn,epsfig,rotate,showpacs]{revtex4}
\usepackage{epsfig}
\usepackage{graphicx}
\usepackage{amsmath}

\newcommand{\beq}{\begin{eqnarray}}
\newcommand{\eeq}{\end{eqnarray}}
\begin{document}
\title{The Born Rule in Quantum and Classical Mechanics}
\author{Paul Brumer and Jiangbin Gong}
\affiliation{Chemical Physics Theory Group \\ and
Center for Quantum Information and Quantum Control\\
University of Toronto, Toronto, M5S 3H6, Canada}
\date{January 17, 2006}
\begin{abstract}
Considerable effort has been devoted to deriving the Born rule 
(e.g. that $|\psi(x)|^2 dx$ is the probability of finding a system,
described by $\psi$, between $x$ and $x + dx$) in
quantum mechanics.  Here we show that the Born rule is not solely
quantum mechanical; rather, it arises naturally in the Hilbert space
formulation of {\it classical} mechanics as well.
These results provide new insights into the nature of the Born rule, and 
impact on its understanding in the framework of quantum mechanics.
\end{abstract}
\pacs{03.65.Sq, 03.65.Ca}
\maketitle

\section{introduction}
The Born rule \cite{born} postulates a connection between deterministic quantum
mechanics in a Hilbert space formalism with probabilistic predictions of
measurement outcomes. It is typically stated \cite{fine} as follows (without
considering degeneracies): if an observable $\hat{O}$, with eigenstates
$\{|O_{i}\rangle\}$ and spectrum $\{O_{i}\}$,  is measured on a system
described by the state vector $|\psi\rangle$, the probability for the
measurement to yield the value $O_{i}$ is given by $|\langle
O_{i}|\psi\rangle|^{2}$. Alternatively, in the density matrix formulation
used below, this rule states that the probability is $Tr\left[\rho_{\psi}
\rho_{O_i}\right]$, where $\rho_{\psi} = |\psi \rangle \langle\psi|$ and
$\rho_{O_i} = |O_i \rangle \langle O_i|$. Most familiar is the textbook example
that the probability of observing a system that is in a state $\psi$
in the coordinate range $x$ to $x + dx$ is given by $|\langle x | 
\psi \rangle|^2 dx.$
Born's rule appears as a
fundamental postulate in quantum mechanics and is thus far in agreement with
experiment. Hence, there is intense interest in providing an
underlying motivation for, or derivation of, this rule.

Gleason's theorem \cite{gleason}, for example,  provides a formal
motivation of the Born rule, but it is a purely mathematical result about
vectors in Hilbert spaces and does not provide insight into the physics of
this postulate. For this reason there have been several
attempts to provide a physical derivation of the Born rule.
For example, Deutsch showed the possibility of deriving the Born rule
from ``the non-probabilistic axioms of quantum theory'' and 
``the non-probabilistic part of classical decision theory''
\cite{deutsch}. Deutsch's approach was 
criticized by Barnum {\it et al.} \cite{barnum2} but was 
recently reinforced by Saunders \cite{saunders}. 
Hanson \cite{hanson} and Wallace \cite{wallace} 
analyzed the connection between possible derivations of the Born rule and
Everett's many worlds interpretation of quantum mechanics.
Zurek recently proposed a significantly new approach,  the so-called
``envariance" approach  \cite{zurek}, for deriving the Born rule from within quantum
mechanics. This approach,
totally different from Deutsch's method, was recently analyzed in 
detail by Schlosshauer and Fine \cite{fine} and by Zurek \cite{zurekpra}.
Zurek's ``envariance"  approach has also been 
analyzed and modified by Barnum \cite{barnum}.
All these studies have attracted considerable interest in deriving 
Born's rule 
by making some basic assumptions about quantum
probabilities or expectation values of observables. 

The Born rule is not expected to violate
any future experiments.  In this sense,  even a strict derivation of the Born rule
will not help predict any experimentally new physics.
However, understanding the origin of
the Born rule is important for isolating
this postulate from other concepts in quantum
mechanics and for understanding what is 
truly unique in quantum mechanics as compared with
the classical physics.
For example, based on the above-mentioned efforts to derive 
the Born rule \cite{deutsch,saunders,zurek,fine,zurekpra,barnum}, 
it seems now clear that
the physical origin of Born's rule is unrelated to the details
(e.g.,  wavefunction collapse) of quantum measurement processes.

The main purpose of this work is to show that Born's rule is not solely
quantum mechanical and that it arises naturally in the Hilbert space
formulation of {\it classical} mechanics.
As such,  the Born rule connecting probabilities with 
eigenvalues and eigenfunctions is not as ``quantum" as
it sounds. Indeed, quantum-classical
correspondence, which played no role in Born's original 
considerations \cite{born,nobel}, can then be arguably regarded 
as an interesting motivation for the Born rule in quantum mechanics.
Similarly, exposing the Born rule in classical mechanics
should stimulate new routes to understanding the physical
origin of this rule in quantum mechanics. In particular, 
new and interesting questions can be asked in connection with 
the previous derivations of the Born rule. 

To demonstrate that the Born rule exists in both quantum and classical
mechanics we: (1) recall that both quantum and classical
mechanics can be formulated in the Hilbert space of density operators
\cite{koopman,prigogine,zwanzig,fano,wilkie,jaffe}, that the quantum and classical systems are 
represented by
vectors $\rho$ and $\rho_c$, respectively, in that Hilbert space, and that
$\rho$ and $\rho_c$ can be expanded in eigenstates of a set of
commuting quantum and classical superoperators, respectively; and show (2) 
that the quantum mechanical Born rule can be expressed in terms of
the expansion coefficient of a given density associated with 
eigendistributions of a set of superoperators in the Hilbert space of density
matrix; and (3) that the classical
interpretation of the phase space representation of $\rho_c$ as a
probability density allows the extraction of Born's rule in classical
mechanics, and gives exactly the same structure as the quantum
mechanical Born rule. 

These results suggest that the quantum mechanical 
Born rule not only applies to cases of large quantum numbers, but
also has a well-defined purely classical limit. 
Hence, independent of other subtle elements of the quantum theory,
the inherent consistency  with the classical Born rule 
for the macroscopic world imposes an important 
condition on any eigenvalue-eigenfunction
based probability rule in quantum mechanics.

\section{The Quantum Mechanical Born Rule in Density Matrix Formalism} 
Consider first quantum mechanics in the Hilbert space of the 
density matrix \cite{fano,wilkie}. Given an
operator $\hat{O}\equiv \hat{K}_{N}$ for a system of $N$ degrees of
freedom, we first consider the (classically) integrable case where 
there exist $N$ independent and commuting observables $\hat{K}_{i},\
i=1, \cdots, N$. Another extreme, the
chaotic case, will be discussed in Sec. IV. For convenience we also assume 
that the $\hat{K}_{i}, \ i=1,2,\cdots, N$,
have a discrete spectrum, but the  central result below
applies to cases with a continuous spectrum as well. The complete set of
commuting superoperators in the quantum Hilbert space can be constructed
as
\begin{eqnarray}
\frac{1}{\hbar}[\hat{K}_{i},\ ],\ \   \frac{1}{2}[\hat{K}_{i},\ ]_{+}, \ \
(i=1,2, \cdots, N),
\end{eqnarray}
where $[,]$ denotes the commutator and $[,]_{+}$ denotes the
anticommutator, i.e., $[\hat{A},
\hat{B}]=\hat{A}\hat{B}-\hat{B}\hat{A}$, $[\hat{A},
\hat{B}]_{+}=\hat{A}\hat{B}+\hat{B}\hat{A}$. The simultaneous
eigendensities  of the complete set of superoperators are denoted
$\rho_{{\bf \alpha},{\bf \beta}}$. That is,
\begin{eqnarray}
\frac{1}{2}[\hat{K}_{i},\rho_{{\bf \alpha}, {\bf \beta}}]_{+} & = &
\alpha_{i}\rho_{{\bf \alpha}, {\bf \beta}}; \nonumber \\
\frac{1}{\hbar}\left[\hat{K}_{i},\rho_{{\bf \alpha}, {\bf \beta}}\right] & = & \beta_{i}\rho_{{\bf \alpha}, {\bf \beta}},
\end{eqnarray}
where ${\alpha}\equiv (\alpha_{1}, \alpha_{2}, \cdots, \alpha_{N})$ is the
collection of eigenvalues associated with 
$\frac{1}{2}[\hat{K}_{i},\  ]_{+}$,
and
${\beta}\equiv (\beta_{1}, \beta_{2}, \cdots, \beta_{N})$ is the
collection of eigenvalues associated with $\frac{1}{\hbar}[\hat{K}_{i},\ ]$.

The state of the quantum system is described by an arbitrary density
matrix ${\rho}$ in the Hilbert space under consideration, and can be expanded
in terms of the basis states $\rho_{{\bf \alpha}, {\bf \beta}}$ as
\begin{eqnarray}
{\rho}=\sum_{{\bf \alpha},{\bf \beta}}D_{{\bf \alpha},{\bf \beta}}
\rho_{{\bf \alpha}, {\bf \beta}},
\label{qex}
\end{eqnarray}
where the sum is over all eigendensities. The sum  
should be understood as an integral if the spectrum is
continuous. Clearly, the expansion coefficients in Eq. (\ref{qex}) are
given by
\begin{eqnarray}
D_{{\bf \alpha},{\bf \beta}}=Tr\left [{\rho}\rho_{{\bf \alpha}, {\bf
\beta}}^{\dagger}\right].
\label{qcoe}
\end{eqnarray}
Equations (\ref{qex}) and (\ref{qcoe}), i.e. the expansion of $\rho$ in
terms of the eigendensities $\rho_{{\bf \alpha},{\bf \beta}}$, is central to
the analysis later below.

Consider now the quantum probability
${\cal P}_Q({\bf K}')$ of finding the quantum observables
$\hat{K}_{i}$ with eigenvalues $K_{i}'$, $i=1,2, \cdots, N$, given that 
the system is in state $\rho$. 
We show here that the Born rule is then equivalent to the statement that
${\cal P}_{Q}({\bf K}')$ {\it must be proportional to the
expansion coefficient $D_{{\bf K}',{\bf 0}}$
of the given density $\rho$ associated with the common
eigendistribution $\rho_{{\bf K}',{\bf 0}}$ of superoperators
$\frac{1}{2}[\hat{K}_{i},\ ]_{+}$
with eigenvalues $K_{i}'$ and of superoperators
$\frac{1}{\hbar}[\hat{K}_{i},\  ]$ with eigenvalue zero}.
To see this, consider first a quantum density for a pure quantum state, e.g., 
$\rho=|\psi\rangle\langle\psi|$ 
(the extension to mixed states is straightforward).
Then the Born rule gives that
\begin{eqnarray}
{\cal P}_Q({\bf K}')
& = & |\langle \psi|{\bf K}'\rangle|^{2}=
 Tr \left[|{\bf K}'\rangle \langle {\bf K}'| |\psi\rangle\langle\psi| \right] \nonumber \\
& = & Tr \left[|{\bf K}'\rangle \langle {\bf K}'| \rho \right] ,
 \label{qborn1}
 \end{eqnarray}
where $|{\bf K}'\rangle$ is a common and normalized eigenfunction of
operators $\hat{K}_{i},\ i=1,2,\cdots, N$.
However,  
\begin{eqnarray}
\frac{1}{2}[\hat{K}_{i}, |{\bf K}'\rangle \langle {\bf K}'|]_{+}  =  K_{i}' |{\bf K}'\rangle \langle {\bf K}'|;
\ [\hat{K}_{i}, |{\bf K}'\rangle \langle {\bf K}'|] = 0,
\end{eqnarray}
so that $|{\bf K}'\rangle \langle {\bf K}'|$ is seen to be the common
eigendistribution of superoperators $\frac{1}{2}[\hat{K}_{i},\ ]_{+}$
with eigenvalues $K_{i}'$ and of superoperators
$\frac{1}{\hbar}[\hat{K}_{i},\  ]$ with eigenvalue zero. That is, 
\begin{eqnarray}
|{\bf K}'\rangle \langle {\bf K}'|=\rho_{{\bf K}',{\bf 0}}.
\label{qborn2}
\end{eqnarray}
Equations (\ref{qcoe}), (\ref{qborn1}), and (\ref{qborn2}) then lead to
\begin{eqnarray}
{\cal P}_{Q}({\bf K}')= D_{{\bf K}',{\bf 0}}.
\label{qeq}
\end{eqnarray}
Equation (\ref{qeq}) is a general restatement of the 
quantum mechanical Born rule
based on the Hilbert space structure of density matrix.

Note that the multidimensional result of Eq. (\ref{qborn1}) has
carefully accounted for possible degeneracies associated with
$K_{N}'$. That is, the total probability of observing $K_{N}'$ would be obtained
by summing ${\cal P}_{Q}({\bf K}')$ with all possible
$K_{i}'$, $i=1,2, \cdots, (N-1)$, a necessary procedure not explicitly
stated in Born's rule. 


\section{The Born Rule in Classical Mechanics}
Consider now classical mechanics. The mechanics has numerous equivalent
formulations, such as Newton's Laws, the Lagrangian or Hamiltonian
formulations, Hamilton-Jacobi theory, etc. The less familiar Hilbert space
formulation of classical mechanics used below was first established by
Koopman \cite{koopman} and subsequently appreciated by, for example,
Prigogine \cite{prigogine}, Zwanzig \cite{zwanzig} and us \cite{wilkie,jaffe}
in some theoretical
considerations. This being the case, the above eigenvalue-eigenfunction
structure is not unique to quantum mechanics, a fact that may not be well
appreciated and that is exploited below.

It is convenient to introduce the classical picture in the phase space
representation, although abstract Hilbert space formulations may be used
as well. Consider then the same case as above. Let the classical limit
of the Wigner-Weyl representation of $\hat{K}_{i}$ be ${K}_{i}({\bf p},
{\bf q})$, where $({\bf p}, {\bf q})$ are momentum and coordinate space
variables. The complete set of commuting superoperators on the classical
Hilbert space is then \cite{wilkie}
\begin{eqnarray}
i\{K_{j}({\bf p}, {\bf q}),\ \},\  K_{j}({\bf p}, {\bf q}), \ \ \ j=1,2, \cdots, N,
\end{eqnarray}
where $K_{j}({\bf p}, {\bf q})$ are multiplicative operators and $\{\, ,
\}$ denotes the classical Poisson bracket. The simultaneous eigendensities of
this complete set of classical operators, denoted $\rho^{c}_{{\bf
\alpha}^{c}, {\bf \beta}^{c}}({\bf p}, {\bf q})$, satisfy:
\begin{eqnarray}
K_{j}({\bf p}, {\bf q})\rho^{c}_{{\bf \alpha}^{c}, {\bf \beta}^{c}}({\bf p}, {\bf q})&=& \alpha_{j}^{c} \rho^{c}_{{\bf \alpha}^{c}, {\bf \beta}^{c}}({\bf p}, {\bf q}); \nonumber \\
i\{K_{j}({\bf p}, {\bf q}),\rho^{c}_{{\bf \alpha}^{c}, {\bf \beta}^{c}}({\bf p}, {\bf q})\}&= &\beta_{j}^{c} \rho^{c}_{{\bf \alpha}^{c}, {\bf \beta}^{c}}({\bf p}, {\bf q}),
\end{eqnarray}
where the notation ${\bf \alpha}^{c}, {\bf \beta}^{c}$, $\alpha_{j}^{c}$,
and $\beta_{j}^{c}$ is introduced in parallel with the quantum case. An
arbitrary classical probability density $\rho_{c}({\bf p}, {\bf q})$ can
be expanded as
\begin{eqnarray}
\rho_{c}({\bf p}, {\bf q})=\sum_{{\bf \alpha}^{c}, {\bf \beta}^{c}} D^{c}_{{\bf \alpha}^{c}, {\bf \beta}^{c}}
\rho^{c}_{{\bf \alpha}^{c}, {\bf \beta}^{c}}({\bf p}, {\bf q}),
\label{classicalsum}
\end{eqnarray}
where
\begin{eqnarray}
D^{c}_{{\bf \alpha}^{c}, {\bf \beta}^{c}} &=& \int d{\bf p} d{\bf q}\
\rho_{c}({\bf p}, {\bf q}) \left[ \rho^{c *}_{{\bf \alpha}^{c}, {\bf
\beta}^{c }}({\bf p}, {\bf q})\right] \nonumber \\
&\equiv& Tr\left[\rho_c \rho^{c \dagger}_{{\bf
\alpha}^{c}, {\bf \beta}^{c}}\right],
\end{eqnarray}
and where the sum in Eq. (\ref{classicalsum}) is over all eigendensities.

Consider now, within this formalism, the probability of finding ${\bf K}$
[with ${\bf K} \equiv (K_{1},K_{2}, \cdots, K_{N})$]
between ${\bf K'}$ and ${\bf K'} + d{\bf K'}$.
To proceed we make a canonical transformation between representations $({\bf p}, {\bf q})$ 
and $({\bf K}, {\bf Q})$, where ${\bf K}$  are the new momentum variables,
and ${\bf Q}\equiv(Q_{1},Q_{2},
\cdots, Q_{N})$ denotes the new position variables conjugate to ${\bf K}$. 
The ${\bf Q}$ can be obtained by
regarding ${\bf p}$ as a function of ${\bf q}$ and ${\bf K}$, 
defining the generating 
function $S({\bf q}, {\bf K})=\int_{{\bf q}_{0}}^{{\bf q}} {\bf p}({\bf q}', {\bf K})
\cdot d{\bf q}'$, and then noting that $Q_{i}=\partial S/\partial
K_{i}$ \cite{rasband}.
In this representation the classical eigendensities
$\rho^{c}_{\alpha^{c}, {\bf \beta}^{c}} $ take a rather simple form,
\begin{eqnarray}
\rho^{c}_{{\bf \alpha}^{c}, {\bf \beta}^{c}}({\bf p}, {\bf q})& =>&  
{\cal R}_{{\bf K}',{\bf \Lambda}}({\bf K}, {\bf Q}) \nonumber \\
&=&
\frac{1}{{(2\pi)}^{N/2}}\delta({\bf K}'-{\bf K})
\exp(i{\bf \Lambda}\cdot{\bf Q}) 
\label{eigendist}
\end{eqnarray}
with eigenvalues \cite{footnote} $\alpha^{c}_{i}  =   K_{i}$ and $\beta^{c}_{i}  = \Lambda_{i}$,
for $i=1,2,\cdots, N$.
The set of eigendistributions
${\cal R}_{{\bf K}',{\bf \Lambda}}$ are complete and orthogonal, i.e.,
\begin{eqnarray}
& & \int d{\bf K}d{\bf \Lambda}\ {\cal R}^{*}_{{\bf K},{\bf \Lambda}}({\bf K}', {\bf Q}')
{\cal R}_{{\bf K},{\bf \Lambda}}({\bf K}'', {\bf Q}'') \nonumber \\
&&=
\delta({\bf K}'-{\bf K}'') \delta({\bf Q}'-{\bf Q }''); \nonumber \\
&& \int d{\bf K} d{\bf Q}\ {\cal R}^{*}_{{\bf K}',{\bf \Lambda}'}({\bf K}, {\bf Q})
{\cal R}_{{\bf K}'',{\bf \Lambda}''}({\bf K}, {\bf Q})\nonumber \\
&&= \delta({\bf \Lambda}'-{\bf \Lambda}'')\delta({\bf K}'-{\bf K }'').
\end{eqnarray}
The ${\cal R}_{{\bf K}',{\bf \Lambda}}$ are ``improper states",
insofar as they contain delta functions. However, this is consistent with
the fact that they are eigendistributions of classical superoperators
that have continuous
spectra. If desired, a Rigged Hilbert space \cite{bohm} can be used to
include these states more formally. 

Given a classical probability density $\rho_{c} ({\bf p}, {\bf q})$ that 
describes the state of the system, we can convert to the ${\bf K}, {\bf Q}$
representation to obtain $\rho_{c}'
({\bf K}, {\bf Q})\equiv \rho_{c}\left[{\bf p}({\bf K}, {\bf Q}), 
{\bf q}({\bf K}, {\bf Q})\right]$.
The probability ${\cal P}_{c}({\bf K}')$  of finding the observables 
between ${\bf K}'$ and ${\bf K}'$+d${\bf K}$ is evidently given by
\begin{eqnarray}
{\cal P}_{c}({\bf K}')=
\left[\int d{\bf K} d{\bf Q}\ \delta({\bf K}'- {\bf K}) 
\rho_{c}'({\bf K},{\bf Q})\right] d{\bf K}.
\label{obvious}
\end{eqnarray}
This result has an enlightening interpretation in the Hilbert space formulation
of classical mechanics, as can be seen by 
independently obtaining Eq. (\ref{obvious}) using this 
approach. To do so, we first expand the given density 
in terms of the basis states ${\cal R}_{{\bf K}',{\bf \Lambda}}$. That is,
\begin{eqnarray}
\rho_{c}'({\bf K}, {\bf Q})
= \int d{\bf K}' d{\bf \Lambda}
D^{c}_{{\bf K}',{\bf \Lambda}}{\cal R}_{{\bf K}',{\bf \Lambda}}({\bf K}, {\bf Q}),
\label{expansion}
\end{eqnarray}
where the expansion coefficients are given by  the overlap integrals
\begin{eqnarray}
D^{c}_{{\bf K'},{\bf \Lambda}}&=&
\int d{\bf K}d{\bf Q}\ \rho_{c}'({\bf K}, {\bf Q})
{\cal R}^{*}_{{\bf K}',{\bf \Lambda}}({\bf K}, {\bf Q}). 
\label{overlapeq}
\end{eqnarray}
Because classical probabilities in ${\bf
K}$, that do  not refer to ${\bf Q}$, are obtained by integrating
over ${\bf Q}$, we obtain
\begin{eqnarray}
{\cal P}_{c}({\bf K}')= \left[\int d{\bf Q}\ \rho_{c}'({\bf K}', {\bf Q})\right] d{\bf K}.
\label{peq1}
\end{eqnarray}
Substituting Eq. (\ref{expansion}) into Eq. (\ref{peq1}) and using
Eq. (\ref{eigendist}) then yields
\begin{eqnarray}
{\cal P}_{c}({\bf K}')
&=& (2\pi)^{N/2}  D^{c}_{{\bf K}',{\bf 0}} d{\bf K}
\label{claborn}
\end{eqnarray}
This result is equivalent to  Eq. (\ref{obvious}), but contains an important
message from the perspective of mechanics in Hilbert space.
That is, Eq. (\ref{claborn})
indicates that, given a classical density $\rho'_{c}$ 
that describes the state, the probability of finding 
the observable $K_{j}$ ($j=1,2,\cdots, N$)
in a regime $({\bf K}', {\bf K}' +d{\bf K})$,
is {\it proportional to the overlap between the given density
$\rho'_{c}$  and the common eigendistribution ${\cal R}_{{\bf 
K}',{\bf 0}}$ of multiplicative
operators $K_{j}$ with
eigenvalues $K_{j}'$ and of operators $i\{ K_{j}, \}$
with eigenvalue zero}. This overlap is $D^{c}_{{\bf K}',{\bf 0}}$,
the expansion coefficient of $\rho'_{c}$ in terms of ${\cal R}_{{\bf K}',{\bf 0}}$.
Significantly,  this connection between classical probabilities and
the overlap between the given classical state density and
a particular set of classical eigendistributions [Eq. (\ref{claborn})]
is the direct analog of the quantum
mechanical Born rule in the density matrix formalism [Eq. (\ref{qeq})].
That is, Eq. (\ref{claborn})  is the Born rule in classical mechanics.


\section{Quantum versus Classical Born rule in Chaotic Cases}
Classical-quantum correspondence of Hilbert space structures in chaotic
cases is more subtle and complicated than in integrable cases
\cite{wilkie}.  Nevertheless, results above indicate that
only a particular
set of eigendensities are relevant in understanding the quantum versus classical
Born rule.  
Indeed, it is straightforward to show that
our previous considerations also apply to cases where 
there do not exist $(N-1)$ observables that
commute with  $K_{N}({\bf p}, {\bf q})$.
For example, consider a chaotic
spectrum case
where $K_{N}({\bf p}, {\bf q})$ does not commute with any
other smooth phase space functions $Z({\bf p}, {\bf q})$, i.e.,
$\{K_{N}({\bf p}, {\bf q}), Z({\bf p}, {\bf q})\}\ne 0 $ always holds. 
Following Ref. \cite{wilkie} we consider 
a function $\tau({\bf p}, {\bf q})$ 
that satisfies $\{\tau({\bf p}, {\bf q}), K_{N}({\bf p},
{\bf q})\}=1 $. Then the classical eigenfunction of the multiplicative
operator $K_{N}({\bf p}, {\bf q})$ and of the operator $i\{ K_{N}({\bf p},
{\bf q}), \ \}$ with eigenvalue $\lambda$ is given by $\xi \delta
[K_{N}'-K_{N}({\bf p}, {\bf q})] \exp [i\lambda \tau({\bf p}, {\bf q})]$,
where $\xi$ is a normalization constant. In particular,  the eigenfunction
with zero eigenvalue for the operator $i\{ K_{N}({\bf p}, {\bf q}), \ \}$
is $\xi\delta [K_{N}'-K_{N}({\bf p}, {\bf q})]$.
This eigenfunction defines a $(2N-1)$-dimensional hypersurface in phase
space (whereas the eigenfunction 
${\cal R}_{{\bf K}',{\bf 0}}$ in the integral case defines an $N$-dimensional
manifold).
The overlap between this
particular eigenfunction $\xi\delta [K_{N}-K_{N}({\bf p}, {\bf q})]$ and  a
given phase space probability density $\rho_{c}({\bf p}, {\bf q})$, i.e.,
\begin{eqnarray}
P_{c}(K_{N}')\equiv \xi \int d{\bf p} d{\bf q}\
\rho_{c}({\bf p}, {\bf q})\delta [K_{N}'-K_{N}({\bf p}, {\bf q})],
\label{chaotic}
\end{eqnarray}
yields the probability $P_{c}(K_{N}')dK_{N}$ of finding $K_{N}({\bf p},
{\bf q})$ lying in the regime $[K_{N}', K_{N}'+dK_{N}]$.
This is again in complete analogy to how 
the quantum probability ${\cal P}_Q(K_{N}')$ of finding the eigenvalue $K_{N}'$
is determined, i.e., ${\cal P}_Q(K_{N}')$ is given by the
overlap between a given quantum density $|\psi\rangle\langle\psi|$ and the
eigenfunction of superoperator  $\frac{1}{2}[\hat{K}_{N},]_{+}$
with eigenvalue $K_{N}'$ and of
superoperator $\frac{1}{\hbar}[\hat{K}_{N},\ ]$ with eigenvalue zero. Such a
quantum eigenfunction is simply given by $|K_{N}'\rangle\langle K_{N}'|$,
where $|K_{N}'\rangle$ is the eigenfunction of $\hat{K}_{N}$ with
eigenvalue $K_{N}'$.  These analyses make it clear that even for chaotic cases, 
the Born rule formulated in terms of eigen-densities in
the associated (classical or quantum) Hilbert space
exists in both quantum and classical mechanics. 


\section{Concluding Remarks}
In summary, with the quantum mechanical Born rule
formulated in the Hilbert space of
density
matrix, we have demonstrated an analogous Born rule in classical
mechanics.
In so doing we never assumed that the
quantum density goes smoothly,
in the classical limit, to a classical density that already
has a clear probabilistic interpretation.  Rather, we have simply assumed that a system
in either quantum or classical mechanics is described by a density
operator in Hilbert space, and that these density operators serve the same
descriptive purpose in both mechanics. This, plus the decomposition of the
operator in terms of eigendistributions of a set of commuting superoperators,
suffice to show that the Born
rule applies in both quantum and classical mechanics. Hence,
the quantum mechanical Born rule appears to be very natural in the light of
quantum-classical correspondence in how Hilbert space
structures embody measured probabilities.  
The recognition that Born's rule is not really a unique quantum element
should complement, as well as impact upon, previous attempts to derive the Born rule
\cite{deutsch,saunders,zurek,fine,zurekpra,barnum}.
Further, it motivates numerous questions, such as, can 
the quantum mechanical Born rule be derived with fewer assumptions
by taking advantage of the classical Born rule as a limit?
How can one reconcile derivations involving purely quantum 
language with the existence of a classical Born rule?, etc. These, and
related issues, are the subject of future work.

{\bf Acknowledgments:}
This work was supported by a grant from the National Science and
Engineering Research Council of Canada. We thank Profs. R. Kapral and S. Whittington,
University of Toronto, for comments on an earlier version of this 
manuscript. 


\end{document}